\documentstyle[preprint,aps,floats,epsfig]{revtex}

\newcommand{\be}{\begin{equation}}
\newcommand{\ee}{\end{equation}}
\newcommand{\bea}{\begin{eqnarray}}
\newcommand{\eea}{\end{eqnarray}}

\begin{document}
\tighten
\rightline{hep-th/0003045}
\vspace*{0.5cm}
\begin{center}
{\Large \bf Gravity and antigravity in a brane world with 
metastable gravitons.}\\
\vskip 5mm
{\large Comment on:  G.~Dvali, G.~Gabadadze and M.~Porrati,
``Metastable gravitons and infinite volume extra 
dimensions,'' hep-th/0002190}\\
{\large and}\\
{\large  C.~Csaki, J.~Erlich, T.~J.~Hollowood, 
``Graviton propagators, brane bending and bending of light in 
theories with quasi-localized gravity,'' hep-th/0003020}
\vspace{0.5cm}
\end{center}

\centerline{\bf Ruth Gregory$^{1}$, 
Valery A. Rubakov$^{2}$ and Sergei M. Sibiryakov$^{2}$} 
\centerline{\small \em ~$^1$ Centre for Particle Theory, Durham University,
South Road, Durham, DH1 3LE, U.K.}
\centerline{\small \em {~}$^2$ Institute for Nuclear Research of 
the Russian Academy of Sciences,}
\centerline{\small \em 60th October Anniversary prospect, 7a, 
Moscow 117312, Russia.}
\date{}
\begin{abstract}
In the framework of a five-dimensional three-brane model with
quasi-localized gravitons we evaluate metric perturbations induced on
the positive tension brane by matter residing thereon. 
We find that at intermediate distances, the effective four-dimensional
theory coincides, up to small corrections, with General Relativity.
This is in accord with Csaki, Erlich and Hollowood and in contrast to
Dvali, Gabadadze and Porrati. We show, however, that at ultra-large
distances this effective four-dimensional theory becomes dramatically
different: conventional tensor gravity changes into scalar
anti-gravity. 
\end{abstract}



\newpage

The papers by Dvali, Gabadadze and Porrati, \cite{Dvali:2000rv},
and  Csaki, Erlich and Hollowood, \cite{cs}, 
address the issue of whether four-dimensional gravity 
is phenomenologically acceptable in a class of brane 
models with infinite extra dimensions in which the five-dimensional
gravitons have a metastable ``bound state'', rather than a genuine
zero mode. A model of this sort has been proposed in 
Refs.~\cite{Charmousis:1999rg,Gregory:2000jc} and is a variation
of the Randall--Sundrum (RS) scenario
for a non-compact fifth dimension \cite{Randall:1999vf}.
The construction with metastable gravitons has been put in a more
general setting in Refs.~\cite{Csaki:2000pp,Dvali:2000rv}. It has been
argued in Ref.~\cite{Dvali:2000rv} that models with metastable
gravitons are not viable: from the four-dimensional point of view,
gravitons are effectively massive and hence appear to suffer from a
van Dam--Veltman--Zakharov \cite{DVZ}\ discontinuity in the propagator in the
massless limit. In particular, it has been claimed \cite{Dvali:2000rv}
that the prediction for the deflection of light by massive bodies
would be considerably different from that of General Relativity.
The issue has recently been analyzed in more detail in Ref.~\cite{cs},
where explicit calculations of four dimensional gravity have been
performed along the lines of Garriga and Tanaka \cite{Garriga:1999yh},
and Giddings, Katz and Randall \cite{Giddings:2000mu}. The outcome
of that analysis is that four-dimensional gravity has been claimed to
be in fact Einsteinian, despite the peculiarity of apparently 
massive gravitons.

In this comment we also apply the Garriga--Tanaka (GT) technique
to obtain effective four-dimensional gravity at the linearized level, 
considering as an example the model of 
Refs.~\cite{Charmousis:1999rg,Gregory:2000jc}. We find that at 
intermediate distances (which should extend from microscopic to
very large scales in a phenomenologically acceptable model)
four dimensional gravity is indeed Einsteinian, in accord with
Ref.~\cite{cs} and in contrast to Ref.~\cite{Dvali:2000rv}. However,
at ultra-large scales we find a new phenomenon: four-dimensional gravity changes
dramatically, becoming {\it scalar anti-gravity} rather than
tensor gravity. This may or may not signal an internal
inconsistency of the models under discussion.

To recapitulate, the set up of
Refs.~\cite{Charmousis:1999rg,Gregory:2000jc}  is as follows.
The model has five dimensions and contains one brane
with tension $\sigma > 0$ and two branes with equal
tensions $-\sigma /2$ placed at equal distances 
to the right and to the left of the positive tension
brane in the fifth direction. There is a reflection symmetry,
$z \to -z$, which enables one to consider explicitly
only the region to the right of the positive tension brane
(hereafter $z$ denotes the fifth coordinate). 
Conventional matter resides on the central positive tension brane. 
The bulk cosmological constant between the branes, $\Lambda$, is
negative, whereas it is
equal to zero to the right of the negative tension brane. With
appropriately tuned $\Lambda$,  there exists a
solution to the five-dimensional 
Einstein equations for which both positive and
negative tension branes are at rest at $z=0$ and $z=z_c$ respectively,
$z_c$ being an arbitrary constant. The metric of this solution is
\be
ds^2=a^2(z)\eta_{\mu\nu}dx^{\mu}dx^{\nu}-dz^2
\label{1}
\ee
where
\be
a(z) = \cases{ e^{-kz} & $0<z<z_c$ \cr
e^{-kz_c}\equiv a_- & $z>z_c$\cr}
\label{2}
\ee
The constant $k$ is related to $\sigma$ and $\Lambda$.
The four-dimensional hypersurfaces
$z=const.$ are flat, the five-dimensional space-time is flat to the
right of the negative-tension brane and anti-de Sitter between the
branes. The spacetime to the left of the positive tension brane is
a mirror image of this set-up.

This background has two different length scales, $k^{-1}$
and
\be
r_c = k^{-1} e^{3kz_c}
\ee
These are assumed to be well separated, $r_c \gg k^{-1}$.
It has been argued in Ref.~\cite{Gregory:2000jc} that the extra dimension
``opens up'' both at short distances, $r \ll k^{-1}$ and
ultra-long ones, $r\gg r_c$.

To find the four-dimensional gravity experienced by matter residing
on the positive tension brane, we follow GT and consider a Gaussian-Normal 
(GN) gauge
\be
g_{zz} = - 1\,\;\;\;\;\; g_{z\mu} = 0
\label{n3*}
\ee
In the bulk, one can further restrict the gauge to be transverse-tracefree
(TTF)
\be
h^{\mu}_{\mu} = h^{\mu}_{\nu,\mu} = 0
\label{n3**}
\ee
Hereafter $h_{\mu\nu}$ are metric perturbarions;
indices are raised and lowered by the four-dimensional
Minkowski metric. The linearized
Einstein equations in the bulk take
one and the same simple form for all components of $h_{\mu\nu}$,
\be
\cases{ h'' - 4k^2 h - {1\over a^2} \Box^{(4)}h=0 & $0<z<z_c$\cr
h'' - {1\over a_-^2}\Box^{(4)}h=0 & $z>z_c$\cr}
\label{n4+}
\ee
It is convenient, however, to formulate
the junction conditions on the positive tension brane in the local GN
frame. In this frame, metric perturbations
$\bar{h}_{\mu\nu}$ are not transverse-tracefree, so the two sets of
perturbations are related in the bulk between the two branes by
a five-dimensional gauge transformation preserving (\ref{n3*}),
\be
   \bar{h}_{\mu\nu} = h_{\mu\nu} +\frac{1}{k}\hat{\xi}^5_{,\mu \nu}
      - 2k a^2 \eta_{\mu\nu} \hat{\xi}^5 + a^2 (\xi_{\mu,\nu} +
       \xi_{\nu,\mu})
\label{n4*}
\ee
where $\xi_{\mu}(x)$ and $\hat{\xi}^5 (x)$ are the gauge parameters.
Notice that if $\xi^5$ is not zero, there is a `shift' in the location of 
the wall relative to an observer at infinity, i.e.\ the wall appears
bent to such an observer (as discussed in 
\cite{Garriga:1999yh,Giddings:2000mu}). Physically, this simply represents 
the fact that the wall GN frame is constructed by integrating normal 
geodesics from the wall, and in the presence of matter these geodesics will 
be distorted, thereby altering the proper distance between the wall and 
infinity. In fact, one finds a similar ``bending'' of the 
equatorial plane in the Schwarzschild spacetime if one tries to impose
a local GN frame away from the horizon.

In the presence of additional matter on the positive tension brane
with energy momentum $T_{\mu\nu}$, the junction conditions on this
brane read
\be
\bar{h}'_{\mu\nu} + 2k\bar{h}_{\mu\nu} =
8\pi G_5 \left( T_{\mu\nu} - 
\frac{1}{3} \eta_{\mu\nu} T_{\lambda}^{\lambda} \right)
\label{n4++}
\ee
where $G_5$ is the five-dimensional gravitational constant.
The solution to equations (\ref{n3**}) -- (\ref{n4++}) has been obtained
by Garriga and Tanaka. They found that  $\hat{\xi}^5$ obeys
\be
\Box^{(4)} \hat{\xi}^5 = -\frac{4\pi}{3} 
G_5  T_{\lambda}^{\lambda}
\label{n5++}
\ee
We will need the expression for the induced metric on the positive tension
brane. Up to terms that can be gauged away on this brane, the induced
metric is \cite{Garriga:1999yh}
\be
\bar{h}_{\mu\nu}(z=0) = h^{(m)}_{\mu\nu}
- 2k \eta_{\mu\nu} \hat{\xi}^5
\label{n5+}
\ee
where
\be
h^{(m)}_{\mu\nu} = 16\pi G_5 \int~dx'~
G_R^{(5)}(x,x';z=z'=0)
\left( T_{\mu\nu} - 
\frac{1}{3} \eta_{\mu\nu} T_{\lambda}^{\lambda} \right) (x')
\label{n5*}
\ee
Here $G_R^{(5)}$ is the retarded Green's function of eq. (\ref{n4+})
with appropriate (source-free) junction conditions on the two branes.
This Green's function is mirror-symmetric and obeys
\be
\left[ \partial_z^2 - 4k^2\theta(z_c - z) -\frac{1}{a^2} \Box^{(4)}
+ 4k \delta (z) -2k \delta (z-z_c) \right]  G_R^{(5)}(x,x';z,z')
= \delta (x-x') \delta (z-z')
\ee

Let us consider the case of the static source first. It has been found in 
Ref.~\cite{Gregory:2000jc} that for $k^{-1} \ll r \ll r_c$, the leading
behavior of the static Green's function (given by
$\int~dt~G_R^{(5)}(z=z'=0)$)
is the same as in the RS model (up to small corrections), 
and corresponds to a $1/r$ potential.
Hence, at intermediate distances the analysis is identical to
GT, and the induced metric is the same as in the linearized
four-dimensional General Relativity. This is in accord with 
Ref.~\cite{cs}.

On the other hand, it follows from Ref.~\cite{Gregory:2000jc}
that at ultra-large distances, $r\gg r_c$, 
the contribution (\ref{n5*}) behaves like $1/r^2$ (the fifth dimension
``opens up''). There remains, however, the second term in
eq.(\ref{n5+}). Since eq.(\ref{n5++}) has a four-dimensional
form, this term gives rise to a $1/r$ potential (missed in 
Ref.~\cite{Gregory:2000jc}) even at ultra-large distances. 
For a point-like static source of unit mass, the corresponding
gravitational potential is
\be
V(r) \equiv \frac{1}{2} \bar{h}_{00}(r)
= + \frac{1}{3} G_4 \frac{1}{r}
\ee
where $G_4 = k G_5$ is the four-dimensional Newton's constant
entering also into the
conventional Newton's law at intermediate distances.
We see that at $r\gg r_c$, four-dimensional gravity is induced by
the trace of energy-momentum tensor and has a repulsive $1/r$ potential.
At ultra-large distances tensor gravity changes to scalar
anti-gravity.

Likewise, the four-dimensional gravitational waves emitted by
non-static sources
are conventional tensor ones at intermediate distances and transform
into scalar waves at ultra-large distances (the relevant distance scale
being different from $r_c$ due to relativistic effects, see
Ref.~\cite{Gregory:2000jc}). Indeed, the first term in eq.\ (\ref{n5+})
dissipates \cite{Gregory:2000jc}, whereas the second term survives, 
again due to the four-dimensional structure of eq.\ (\ref{n5++}).

These two cases illustrate the general property of eq.\ (\ref{n5+}):
the first term becomes irrelevant at ultra-large distances (the
physical reason being the metastability of the five-dimensional graviton
bound state), so the four-dimensional
gravity (in effect, anti-gravity) is entirely
due to the second, scalar term. 

This bizarre feature of models with a metastable graviton bound state
obviously deserves further investigation.
In particular, it will be interesting to identify the four-dimensional
massless scalar mode, which is present at ultra-large distances,
among the free sourceless perturbations. This mode is unlikely to
be the radion \cite{Csaki:1999mp,Goldberger:1999un}, 
studied in this model in  Ref.~\cite{Charmousis:1999rg}: the radion
would show up at intermediate distances, as well as at ultra-large 
ones\footnote{The radion presumably couples exponentially weakly to
the matter on the positive tension brane, as it does 
\cite{Garriga:1999yh,Charmousis:1999rg} in the two-brane  model
of Randall and Sundrum
\cite{Randall1}. This may be the reason why the radion effects have
not been revealed by the analyses made in 
Ref.~\cite{cs} and this note.}; 
furthermore, the experience
\cite{Tanaka:2000er} with models where the distance between the branes is
stabilized \cite{Goldberger:1999uk} suggests that the massless 
four-dimensional mode parametrized
by $\hat{\xi}^5$ exists even if the radion is made massive.

More importantly, one would like to understand whether anti-gravity
at ultra-large distances is a signal of an intrinsic inconsistency
of this class of models, or simply a signal that physics is intrinsically
five-dimensional at these scales. In four dimensions, scalar antigravity
requires either negative kinetic and gradient energy or a ghost.
Whether or not a similar 
feature is inherent in models with extra dimensions 
remains an open question. If it is, there still
would remain a possibility that fields with negative energy
might be acceptable, as their effect might show up at
ultra-large distances only.

We note finally, that anti-gravity may not be a special feature
of models with quasi-localized gravitons. 
It is also possible that this phenomenon may
be present in models of the type suggested by Kogan et.\ al.\ 
\cite{Kogan:1999wc}, where some Kaluza--Klein graviton excitations
are extremely light. The same question about internal consistency
then would apply to these models as well.

\vskip 1cm

We would like to thank 
Sergei Dubovsky, Dmitry Gorbunov,
Maxim Libanov and Sergei Troitsky
for useful discussions.  We are indebted
to C.~Csaki, J.~Erlich and T.~J.~Hollowood for 
sending their paper \cite{cs} prior to publication.
R.G.\ was supported in part by the Royal Society, and V.R.\ and S.S.\ by
the Russian Foundation for Basic Research, grant 990218410.

\end{document}